\documentclass[journal,transmag,a4paper]{IEEEtran}
\usepackage{hyperref,color}
\usepackage{amssymb,amsmath,amsfonts}
\usepackage{graphicx,epsfig,subcaption,adjustbox}
\usepackage{xr}
\usepackage[utf8]{inputenc}
\usepackage[bottom=2.3cm,top=2cm,left=1.5cm,right=1.5cm]{geometry}
\usepackage[font=small]{caption}

\usepackage{eso-pic}

\AddToShipoutPicture*{\footnotesize\sffamily\raisebox{0.8cm}{\hspace{1.5cm}\fbox{\parbox{\textwidth}{\copyright~2017
IEEE. Personal use of this material is permitted. Permission from IEEE
must be obtained for all other uses, in any current or future media,
including reprinting/republishing this material for advertising or
promotional purposes, creating new collective works, for resale or
redistribution to servers or lists, or reuse of any copyrighted
component of this work in other works.}}}}

\newcommand{\EqMagVecPotential}{\nabla\times\vec{A}=\vec{B}}

\newcommand{\EqMagVecPotQuasistaticFormulation}{\nabla\times(\nu\ \nabla\times\vec{A}) = \vec{J}_{\text{s}} + \sigma \partial_t{\vec{A}} + \nabla\times\vec{M}_\mathrm{dyn}}

\newcommand{\EqThBalance}{{C_{{p}}}{\partial_t }{T} + \triangledown\cdot\vec{q} = Q}

\newcommand{\EqHeatSources}{Q = Q_\text{dyn}+Q_\text{eddy}+Q_\text{Ohm}}

\newcommand{\EqConstitutiveLaw}{\vec{B} = \nu^{-1}(\vec{H}+\vec{M}_\mathrm{dyn})}

\newcommand{\dBdt}{{\partial_t }\vec{B}}

\newcommand{\EqRhoHalfTurn}{\sigma^{-1}_{\text{ht}}=\frac{\xi_{\text{ht}}}{\kappa_{\text{ht}}f_{\text{Cu}}}\sigma^{-1}_{\text{Cu}}}

\newcommand{\EqRichards}{\xi_{\text{ht}} =\frac{1}{1+e^{-r\left({\frac{|\vec{J}_{\text{s}}|-J_{\text{c,ht}}}{J_{\text{c,ht}}}}-c\right)}}}

\begin{document}


\title{A 2-D Finite-Element Model for Electro-Thermal Transients in Accelerator Magnets}

	\author{
    \centering
		\IEEEauthorblockN{
			L. Bortot\IEEEauthorrefmark{1}, 
            B. Auchmann\IEEEauthorrefmark{1,2},
            I. Cortes Garcia\IEEEauthorrefmark{4},
            A.M. Fernandez Navarro\IEEEauthorrefmark{1},
			M. Maciejewski\IEEEauthorrefmark{1,3},  
		}
        \IEEEauthorblockN{
        	M. Prioli\IEEEauthorrefmark{1}, 
            S. Schöps\IEEEauthorrefmark{4}, 
            and A.P. Verweij\IEEEauthorrefmark{1}
        }            
		\IEEEauthorblockA{\IEEEauthorrefmark{1} 
        		{CERN,  Geneva, Switzerland, E-mail: lorenzo.bortot@cern.ch}}
        \IEEEauthorblockA{\IEEEauthorrefmark{2} 
        		{Paul Scherrer Institut, Villigen, Switzerland}}        
		\IEEEauthorblockA{\IEEEauthorrefmark{3} 
        		{Łódź University of Technology, Łódź, Poland}}
		\IEEEauthorblockA{\IEEEauthorrefmark{4} 
        		{Technische Universität Darmstadt, Darmstadt, Germany}}}


\IEEEtitleabstractindextext{%
\begin{abstract}
Superconducting accelerator magnets require sophisticated monitoring and means of protection due to the large energy stored in the magnetic field. Numerical simulations play a crucial role in understanding transient phenomena occurring within the magnet, and can, therefore, help to prevent disruptive consequences. We present a 2-D FEM model for the simulation of electro-thermal transients occurring in superconducting accelerator magnets. The magnetoquasistatic problem is solved with a modified magnetic vector potential formulation, where the cable eddy currents are resolved in terms of their equivalent magnetization. The heat balance equation is then investigated, and the relevant heat sources are discussed. The model implements a two-port component interface and is resolved, as part of an electrical circuit, in a cooperative simulation scheme with a lumped-parameter network.
\end{abstract}
	
}
	   

\maketitle


\section{Introduction} \label{Introduction}
\IEEEPARstart{C}{ircular} accelerators for high-energy particle physics require intense magnetic fields to control the trajectories of the particle beams. These fields are generated by means of high-field superconducting magnets, which are electrically connected in series, and operated in circuits that contain up to hundreds of elements. It is of paramount importance to ensure a safe management of the stored energy which, if released in an uncontrolled way, could compromise the integrity of the superconducting circuit. This is critical in case of an event such as a quench~\cite{wilson1983superconducting}, where the energy is released as Ohmic losses. Quenches cannot always be prevented, and must be considered among the possible operational scenarios. Generally, dedicated quench detection and protection systems are in place to quickly discharge the stored energy, in order to avoid overheating of the coil. These systems influence both, the quench evolution in the coil and the electrical transient in the rest of the circuit. A careful analysis of the electro-thermal transient is fundamental for the design of both, the magnet and the quench protection system, and for the safe operation of the circuit. 

We present a finite-element electro-thermal 2-D model of a superconducting magnet,
resolved at the scale of half-turns over which material properties and physics laws are homogenized.  The model accounts for the non-linear temperature- and field-dependent material properties and for the induced eddy-currents in the cable, as well as in the coil’s copper wedges. 
The model consistently resolves the field formulation in presence of iron saturation~\cite{auchmann2008calculation}, and dynamic effects in both the coil ~\cite{ravaioli2016lumped} and the wedges. In particular, the inter-strand eddy currents formulation follows~\cite{de2004finite}, with an extra averaging operator that is discussed in~\ref{Electrodynamics}.The thermal formulation covers the coil assembly, including structural elements such as insulation foils and wedges, to account for the turn-to-turn and layer-to-layer heat propagation~\cite{Chiesa:2001zm,Novitski:2007lq}. The model has been developed as a modular component of a wider numerical architecture, which implements the concept of cooperative simulation. The aim of the architecture is to resolve the electro-dynamic coupling between the magnet, the protection systems, and the remaining network, leading to consistent simulations.

In this paper, we discuss in detail the electro-thermal formulation, and how it is implemented in the model. Hence we provide a two-port component interface, which is used to co-simulate the model with an external electrical circuit.


\section{Method} \label{method}
\begin{figure*}[tb]
\centering
   \begin{subfigure}[b]{0.26\textwidth}
   		\centering
   		\includegraphics[width=4.0cm]{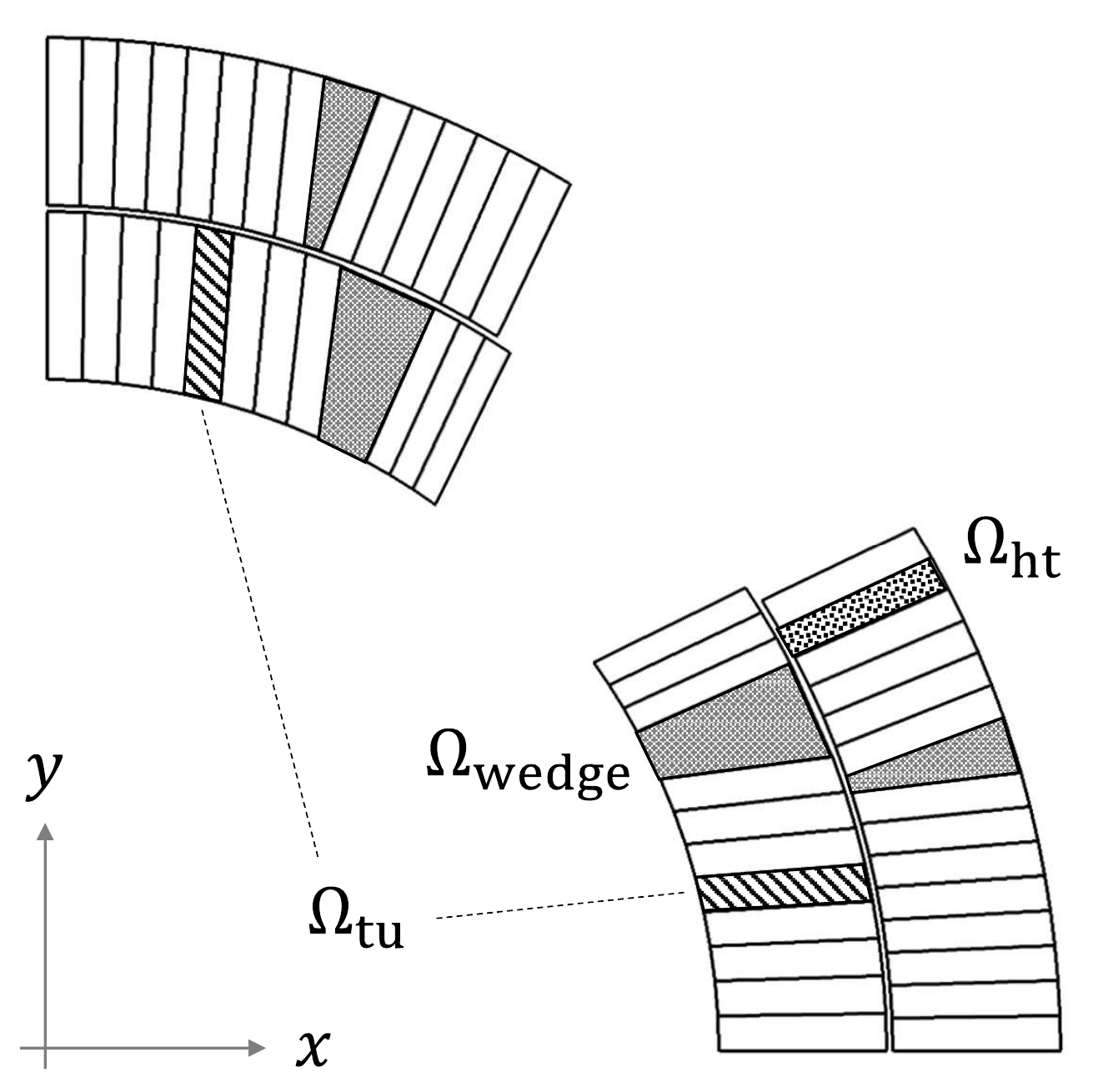}
   		\caption{}
   		\label{MagXsection} 
\end{subfigure}
\begin{subfigure}[b]{0.15\textwidth}
   		\centering
   		\includegraphics[width=2.5cm]{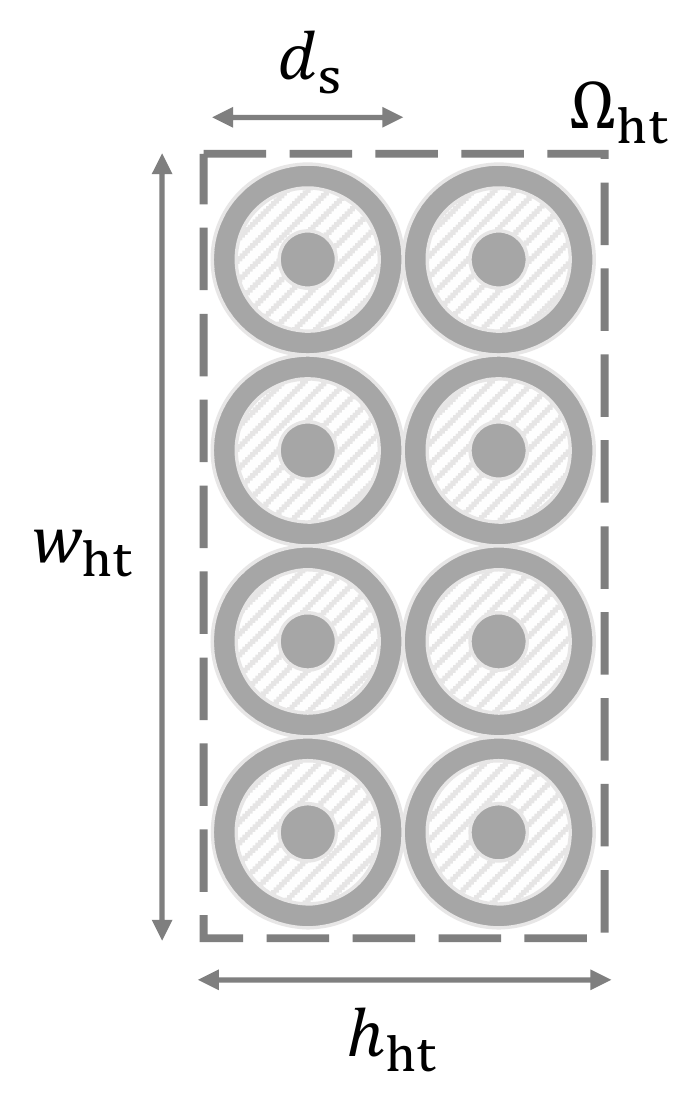}
   		\caption{}
   		\label{TurnHomogenization}
\end{subfigure}
\begin{subfigure}[b]{0.25\textwidth}
   		\centering
   		\includegraphics[width=3.7cm]{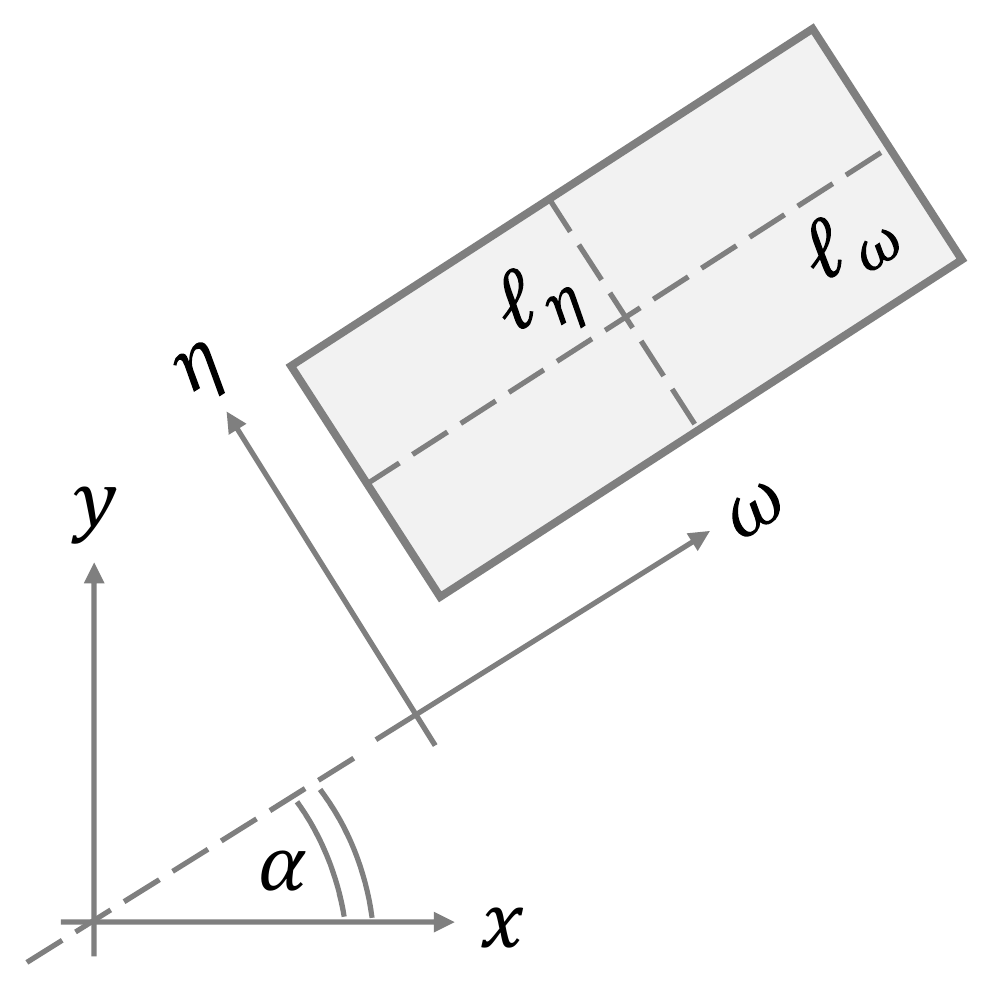}
   		\caption{}
   		\label{TurnReferenceSystem}
\end{subfigure}
\begin{subfigure}[b]{0.32\textwidth}
   		\centering
   		\includegraphics[width=6.0cm]{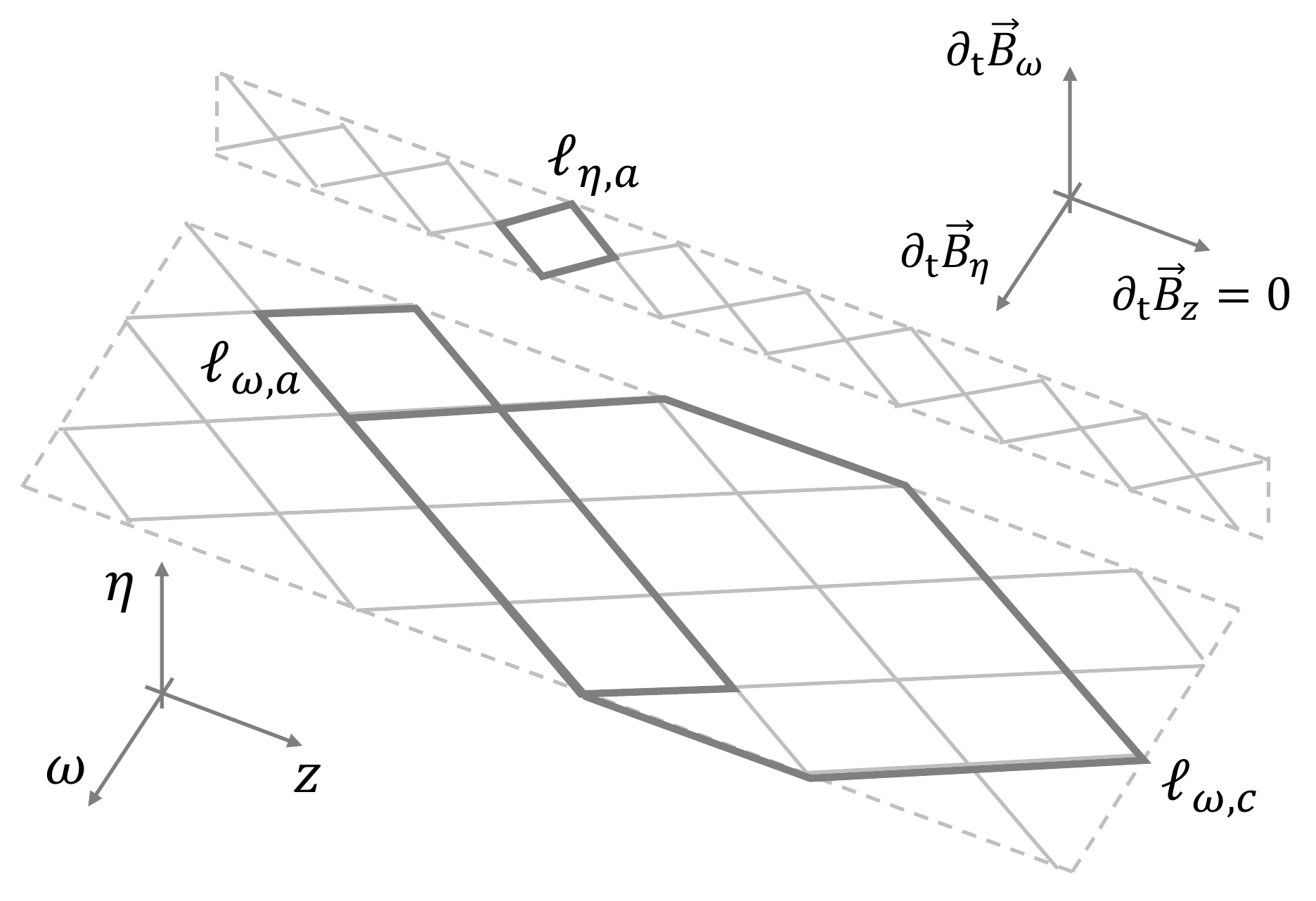}
   		\caption{}
   		\label{RutherfordPattern}
\end{subfigure}
\caption{(a) Single-quadrant cross section of the quadrupole magnet's coil. (b) Actual and homogenized cable's cross section; the light- and dark-grey domains refer respectively to the superconducting and the copper domains; the remaining white domain represents the cable's voids, here considered as filled with epoxy resin. (c) Definition of a $(\eta,\omega)$ local reference frame, as function of $\alpha$. (d) The cable's characteristic strand loops with their associated magnetic flux density components, in a local frame.}
\label{allAboutGeometry} 
\end{figure*}

The electro-thermal model combines the magnetic vector potential formulation $\EqMagVecPotential$ with the heat balance equation.
The magnetoquasistatic field solution, fixed with the Coulomb gauge and driven by the current density source $\vec{J}_{\mathrm{s}}$, determines the magnet's electrodynamics and the related thermal losses. The eddy-currents' term $\sigma\partial_t{\vec{A}}$, proportional to the  conductivity $\sigma$, is replaced for the cable by an equivalent magnetization factor $\vec{M}_\mathrm{cc}$~\cite{wilson1983superconducting,verweij1995electrodynamics} proportional to the time derivative of the magnetic flux density ${\partial_t }\vec{B}$ via an equivalent time constant $\tau_{\text{loop}}$.
The cable's persistent magnetization $\vec{M}_\text{pers}$~\cite{wilson1983superconducting,russenschuck2011field} is considered in $\vec{M}_\mathrm{dyn}=\vec{M}_\text{cc}+\vec{M}_\text{pers}$, leading the dynamic effects to be consistently included in the constitutive law $\EqConstitutiveLaw$, where $\vec{H}$ represents the magnetic field and $\nu^{-1}$ is the magnetic permeability. The temperature field $T$ is determined by the balance of the heat $\rho{C}_{{p}}{\partial_t }{T}$ stored in the system, the heat flux $\triangledown\cdot\vec{q}$, and the heat source $Q$. The contributions to $Q$ stem from dynamic losses $Q_\text{dyn}$ and Ohmic losses $Q_\text{Ohm}$ in the coil, and eddy current losses $Q_\text{eddy}$ in the wedges. The proposed formulation leads, on the 2-D domain $\Omega$, to the following coupled equations, where $\vec{A}=(0,0,A_z)$
\begin{equation}
	\begin{aligned}
			&\EqMagVecPotQuasistaticFormulation,\\
			&\EqThBalance,\\
			&\EqHeatSources.
	\end{aligned}
	\label{SolutionSystem}
\end{equation}
The boundary $\partial{\Omega}$ is linked to $\Omega$ through a layer of infinite domain elements~\cite{beer1981infinite}. Let $\vec{n}$ be the outward pointing vector; if no symmetry is exploited, the Dirichlet boundary condition $\vec{n}\times\vec{A}=0$ is imposed. The model is driven by an external current $I_{\text{s}}$ in order to be coupled with the co-simulation algorithm. The current is distributed over the $n_\text{ht}$ half-turns as $\vec{J}_{\text{s}}=\vec{\chi} I_{\text{s}}$~\cite{leonard1988voltage}, through the density function $\vec{\chi}=\sum_{i=1}^{n_\text{ht}}\vec{\chi}_{\text{ht},i}$. 

To ensure flexibility for the FEM approach, the  model is created by a Java routine that transforms a user's input into a distributed model, relying on the Application Programming Interface (API) provided by $\text{COMSOL}^{\circledR}$. 


\subsection{Homogenization of the cable's geometry} \label{Homogenization}
The magnet coil's discretization resolves the scale of half-turns (ht), which are paired in turns (tu) to form closed loops at infinity, as in Fig.~\ref{MagXsection}. Each half-turn $\Omega_{\text{ht}}$ contains $n_{\text{s}}$ strands of $d_{\text{s}}$ diameter (see Fig.~\ref{TurnHomogenization}), made of a composite of superconducting material and copper whose relative volumetric fractions are defined as $f_{\text{sc}}$ and $f_{\text{Cu}}$. In the model, the generic half-turn's surface $|\Omega_{\text{st}}|=n_{\text{s}}\pi d_{\text{s}}^2/4$ is approximated with a polygon, introducing a discretization error which is compensated through  a suitable homogenization density factor $\kappa_{\text{ht}}={|\Omega_\text{st}|}/{|\Omega_\text{ht}|}$. For later use, we define for each half-turn their tilting angle $\alpha$ and two lines ${\ell_{\omega}}$, ${\ell_{\eta}}$ parallel respectively to the wide and narrow edges (see Fig.~\ref{TurnReferenceSystem}). 


\subsection{Electrodynamics} \label{Electrodynamics}
\begin{figure}[tb]
  \centering
	\includegraphics[width=6.5cm]{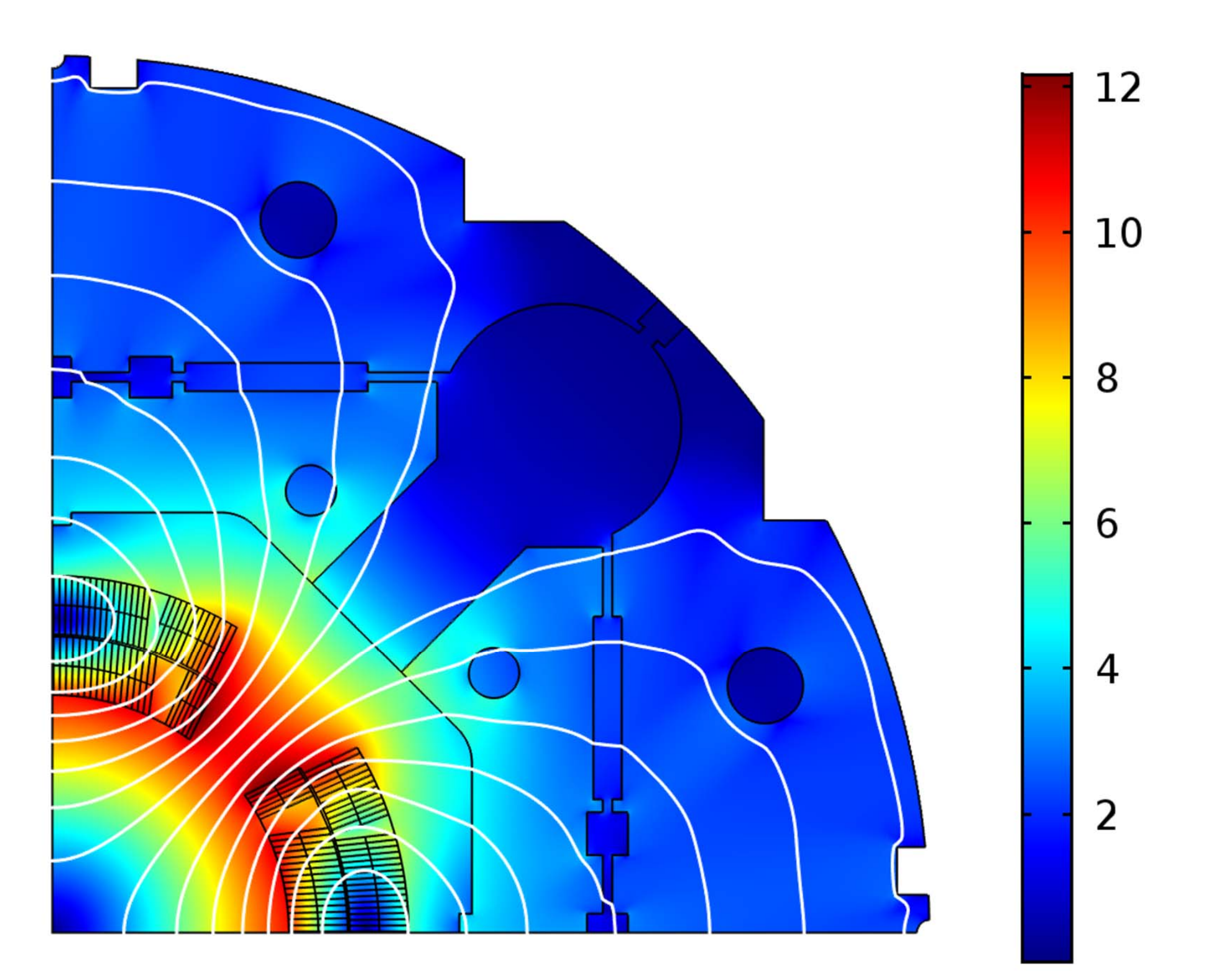}
	\caption{Quadrupole magnet at nominal field, in T.}
	\label{MQXF_Bnom}
\end{figure}

\begin{figure}[tb]
  \centering
	\includegraphics[width=6.5cm]{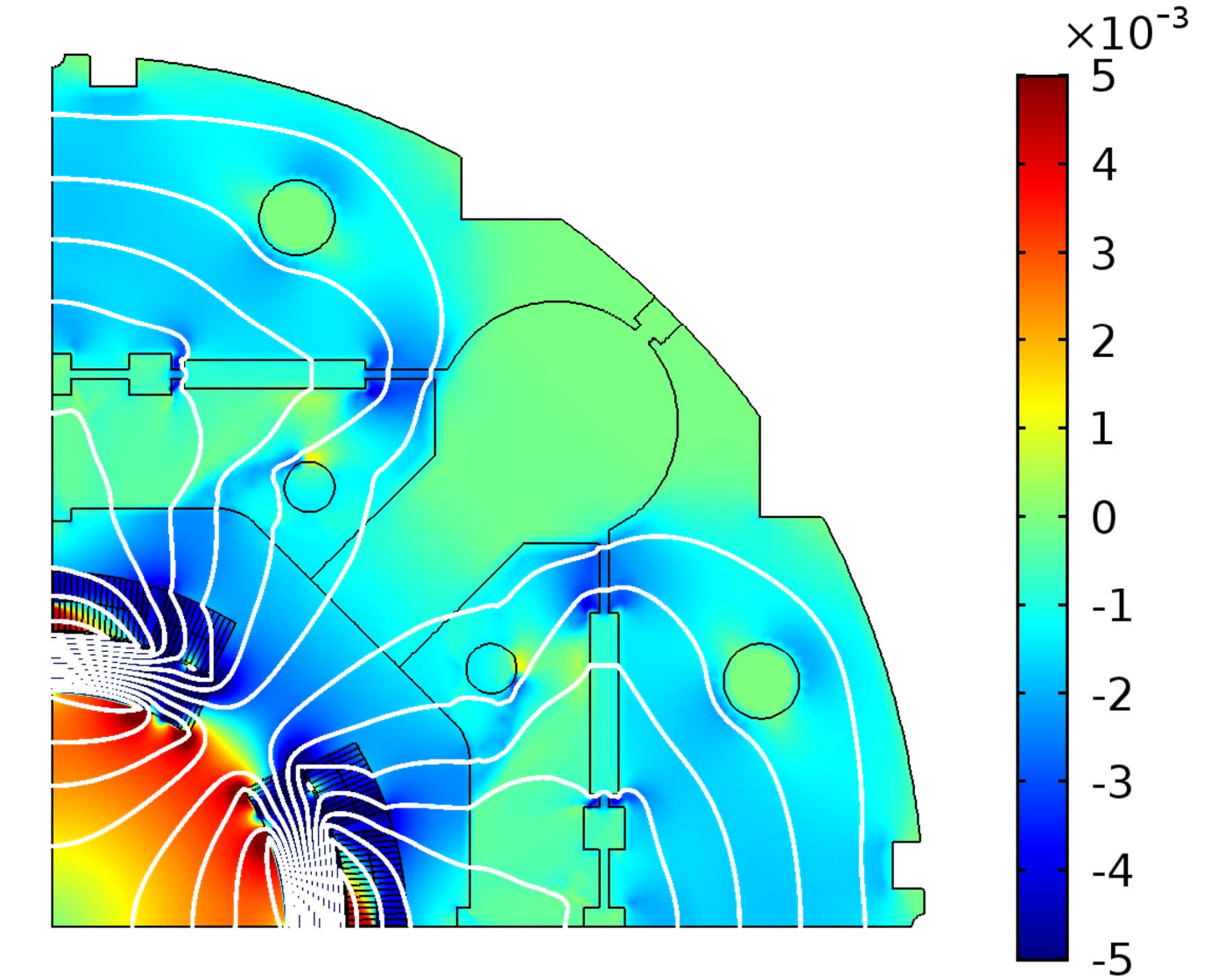}
	\caption{IFCC and ISCC magnetic flux density in T, at $5\text{kA}$, during a linear ramp-up of $100\text{A/s}$, calculated as a difference of two solutions.}
	\label{D1_1q_IFCC_ISCC}
\end{figure}

The superconducting coil features both, inter-filament (IFCC) and inter-strand (ISCC) coupling currents, accounted in the field solution (see Fig.~\ref{MQXF_Bnom}) through their equivalent magnetization, as $\vec{M_\text{cc}}=\vec{M}_\text{IFCC}+\vec{M}_\text{ISCC}$ (see Fig.~\ref{D1_1q_IFCC_ISCC}). This avoids to resolve the coil domain at the micrometric scale of the cable's filamentary structure: the equivalent magnetization combines the laws of Faraday and Ampère-Maxwell, assuming an a-priori knowledge of the currents' loops. 

The IFCC's magnetization always counteracts ${\partial_t }\vec{B}$ in a 2-D domain. It features the time constant $\tau_{\text{IFCC}}$ which depends on the magneto-resistivity of the strand's copper matrix, and on the filament's diameter and twist pitch~\cite{verweij1995electrodynamics,morgan1970theoretical},
\begin{equation}
	\vec{M}_\text{IFCC} = -\kappa_\text{ht}\,\nu\,\tau_{\text{IFCC}}\,{\partial_t }\vec{B}.
	\label{Eq_M_IFCC}
\end{equation}

The ISCC's magnetization features three distinct contributions~\cite{wilson27dipole}, which reflect the three characteristic strand loops $\ell_{\mathrm{\omega,c}}$, $\ell_{\mathrm{\omega,a}}$ and $\ell_{\mathrm{\eta,a}}$ associated to the ISCCs paths in a fully transposed cable (see Fig.~\ref{RutherfordPattern}). Each of the three ISCCs contributions is homogenized, then expressed per cable's unit length~\cite{verweij1995electrodynamics} and linked to an equivalent time constant ($\tau_{\omega,c}, \tau_{\omega,a} ,\tau_{\eta,c}$). The original loops are replaced by the  equivalent loops ${\ell_{\omega}}$ and ${\ell_{\eta}}$, orthogonal respectively to the cable's wide and narrow edges. The tensor of the loops' equivalent time constants reads, in a local reference frame ($\omega,\eta$), as
\begin{equation}
\tau_{\text{ISCC}} = 
	\begin{bmatrix}
		\tau_{\omega,c} + \tau_{\omega,a} & 0
		\\0 & \tau_{\eta,a}
		\end{bmatrix}.
\label{eqMagBasics}
\end{equation}
Due to the formulation's dependency on the turns' orientations, it is suitable to introduce the rotation matrix ${R(\alpha)}$, positive definite for a counterclockwise rotation of the 2-D Euclidean space. It is also convenient to define the $P(\ell_{\mathrm{\omega,\eta}})$ operator, which provides the average of the normal components of a given vectorial field over $\ell_{\mathrm{\omega}}$ and $\ell_{\mathrm{\eta}}$. Hence, we define $\tau_{\text{ISCC},\alpha} = {R(-\alpha)}\tau_{\text{ISCC}}{R(\alpha)}$ and $\dBdt_{\alpha} = {R(-\alpha)}{P(\ell_{\mathrm{\omega,\eta}})}{R(\alpha)}\dBdt$, leading to
\begin{equation}
	\vec{M}_{\text{ISCC}} =-\kappa_\mathrm{ht}\,\nu\,\tau_{\text{ISCC},\alpha}\,\dBdt_{\alpha}.
	\label{Eq_M_ISCC}
\end{equation}

The persistent magnetization term $\vec{M}_\text{pers}$ accounts for the cable's persistent eddy currents circulating in the superconducting filaments and it is implemented as in~\cite{wilson1983superconducting,russenschuck2011field}, according to the Critical State Model~\cite{bean1964magnetization}, neglecting the hysteretic behavior. The critical current density $J_{\text{c}}(B,T)$~\cite{wilson1983superconducting}, implemented as~\cite{bottura2009j_}, is homogenized as $J_{\text{c,ht}}=\kappa_{\text{ht}}f_{\text{sc}}J_{\text{c}}$, leading to
\begin{equation}
	\vec{M}_{\text{pers}}=-\frac{2}{3\pi}d_{\text{f}}J_{\text{c,ht}}\left(1-\frac{|\vec{J}_\text{s}|^2}{J_{\text{c,ht}}^2}\right)\vec{u}_{B}
	\label{Eq_M_PersMag}
\end{equation}
where $d_{\text{f}}$ represents the superconducting filaments' diameter, and $\vec{u}_{B}={\vec{B}}/{|\vec{B}|}$ is the magnetic flux density versor. $\vec{M}_\text{pers}$ decays rapidly with the increase of $I_\mathrm{s}$, being negligible during high-field operations.

The magnet's structural elements are subjected to eddy currents, which are negligible in the laminated iron yoke and steel collar, but relevant in the copper wedges in the coil assembly. The wedges do not form  closed loops at infinity and do not have external leads, so that we require $\int_{\Omega_\text{wedge}}{(\vec{J}_{\text{eddy}}\cdot\vec{n}})\ \mathrm{d}\Omega=0$.


\subsection{Heat Balance} \label{Heat Balance}

The thermal diffusivity of copper dominates that of the insulation, at cryogenic temperature ($10\gg0.05~\text{mm}^{2}/\text{ms}$,~\cite{manfreda2011review}). Therefore, we consider $\Omega_{\text{ht}}$ as an isothermal domain, over which the material properties are averaged. 
The heat diffusion term implements the Fourier Law $\vec{q}=-k\triangledown{T}$ and it accounts for the heat exchanged through both, the insulation layer of each half-turn, and the insulation foil between the inner and outer stack of cables. Due to the aspect ratio of the half-turns' insulation layer ($>{100}$ on the turn's wide side), $k$ is redefined as $k_\mathrm{ht}$, symmetric and anisotropic, to neglect the tangential component of the heat flux. In the implementation, an in-built feature called \textit{thin layer} avoids to mesh the half-turn's insulation layer, while the insulation foil is explicitly meshed. The half-turn domains implement a specific heat capacity ${\rho}C_{p,\text{ht}}$ which  accounts for the superconducting material, the copper stabilizer, the external insulation and for the filling material which  impregnates the cable's voids (see Fig.~\ref{TurnHomogenization}). Each material contributes with its volumetric fraction $f_{i}$ to ${\rho}C_{p,\mathrm{ht}} = \kappa_\mathrm{ht}\sum_{i}^{n}f_{i}\rho_{i}C_{p,i}$.

The heat source $Q$ acting on the coil assembly features three main contributions. The specific losses $Q_\mathrm{eddy}$ associated to the eddy currents in the wedges are proportional to the wedges conductivity $\sigma_{\text{wedge}}$, as $Q_\mathrm{eddy}={\vec{J}_\text{eddy}}\cdot\sigma^{-1}_{\text{wedge}}{\vec{J}_\text{eddy}}$.
The specific losses $Q_\text{dyn}$, deposited by the dynamic effects in the coil, are evaluated from the associated variation of the magnetic energy density function~\cite{sorbi2016magnetization}. Losses are determined from
\eqref{Eq_M_IFCC}, \eqref{Eq_M_ISCC}, \eqref{Eq_M_PersMag} as
$Q_\text{dyn}=\vec{M}_{\text{IFCC}}\cdot\dBdt+\vec{M}_{\text{ISCC}}\cdot\dBdt_{\alpha}+\vec{M}_{\text{pers}}\cdot\dBdt$.
The contribution $Q_\text{Ohm}$ related to the Ohmic losses appears when and where the cable's working point moves out of the superconducting material's critical surface~\cite{wilson1983superconducting}, losing the superconducting state. We model the quench state-transition through a suitable transition-variable $\xi_{\text{ht}}$.
Joule losses are accounted as $Q_\mathrm{Ohm}={\vec{J}_\text{s}}\cdot\sigma^{-1}_{\text{ht}}{\vec{J}_\text{s}}$, where the equivalent half-turn resistivity $\sigma^{-1}_{\text{ht}}$ is calculated using a  reformulation of the Stekly approximation~\cite{stekly1965stable} for the current sharing regime~\cite{walters1975design}, simplifying the extraction of the coil's equivalent resistance(Sec.~\ref{Interface}). In detail $\sigma^{-1}_{\text{ht}}$ is scaled by the $C^{1}$ class logistic function~\cite{richards1959flexible} $\xi_{\text{ht}}$ where the tuning coefficients are chosen as $c=0.5$, $r=10$, leading to 
\begin{equation}
  \EqRhoHalfTurn,\\\ \EqRichards,
 \label{rhoEqStekly}
\end{equation}

Alternatively to \eqref{rhoEqStekly}, one can directly fit the $E=f(J)$ with a logistic function, ensuring a finite amount of energy with respect to the power law~\cite{russenschuck2011field}. Nevertheless, in this case the calculation of $\sigma^{-1}_{\text{ht}}$ adds an implicit algebraic equation so that \eqref{rhoEqStekly} has been found to be more efficient.


\subsection{Two-ports component interface} \label{Interface}
From the solution we extract two quantities per coil's unit length: the total resistance $R_\mathrm{c}$ and the inductive voltage $U_\mathrm{c}$, calculated as the time derivative of the coil's linked flux. If the average of the vector potential $\vec{A}_{\text{ht}}$ over a generic half-turn $\Omega_{\text{ht}}$ is introduced, the coil's equivalent parameters read
\begin{equation}
R_{\text{c}} = \sum_{i=1}^{n_\text{ht}}(\sigma^{-1}_{\text{ht},i}\ \Omega_{\text{ht},i}), \ \ \ U_\mathrm{c} =\sum_{i=1}^{n_\text{ht}}\left(\vec{\chi}_{\text{ht},i}\cdot\partial_{t}\vec{A}_{\text{ht},{i}}\right).
\label{}
\end{equation}
Given  the turns' electrical connection order, the definition of the resistive and inductive voltage per turn allows to evaluate the voltage-to-ground distribution along the magnet's coil. 


\section{Results} \label{Result}
\begin{figure}[tb]
\centering
   \begin{subfigure}[b]{0.23\textwidth}
   		\centering
   		\includegraphics[width=4.4cm]{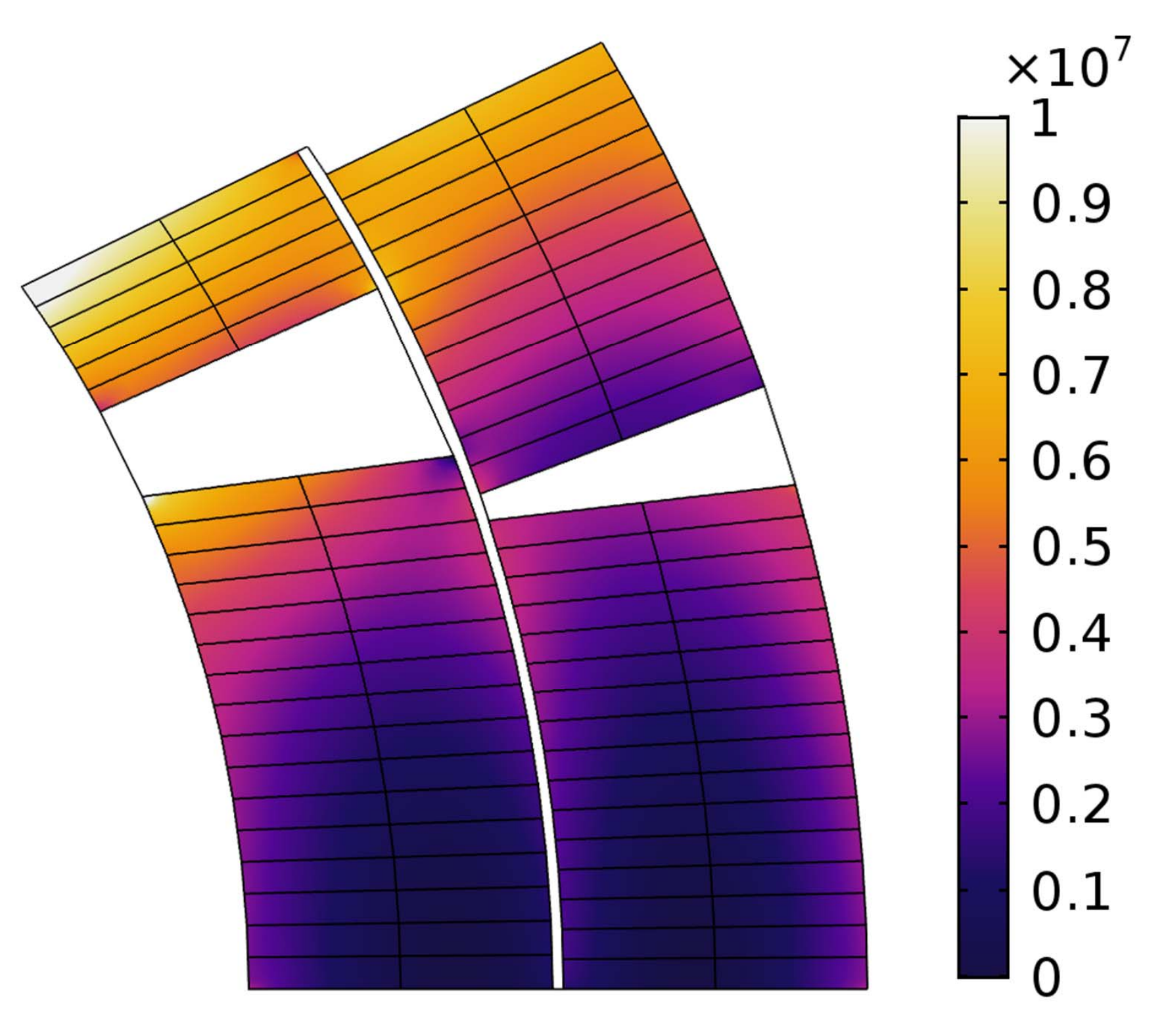}
   		\caption{}
   		\label{fig:No1} 
\end{subfigure}
   \begin{subfigure}[b]{0.23\textwidth}
   		\centering
   		\includegraphics[width=4.4cm]{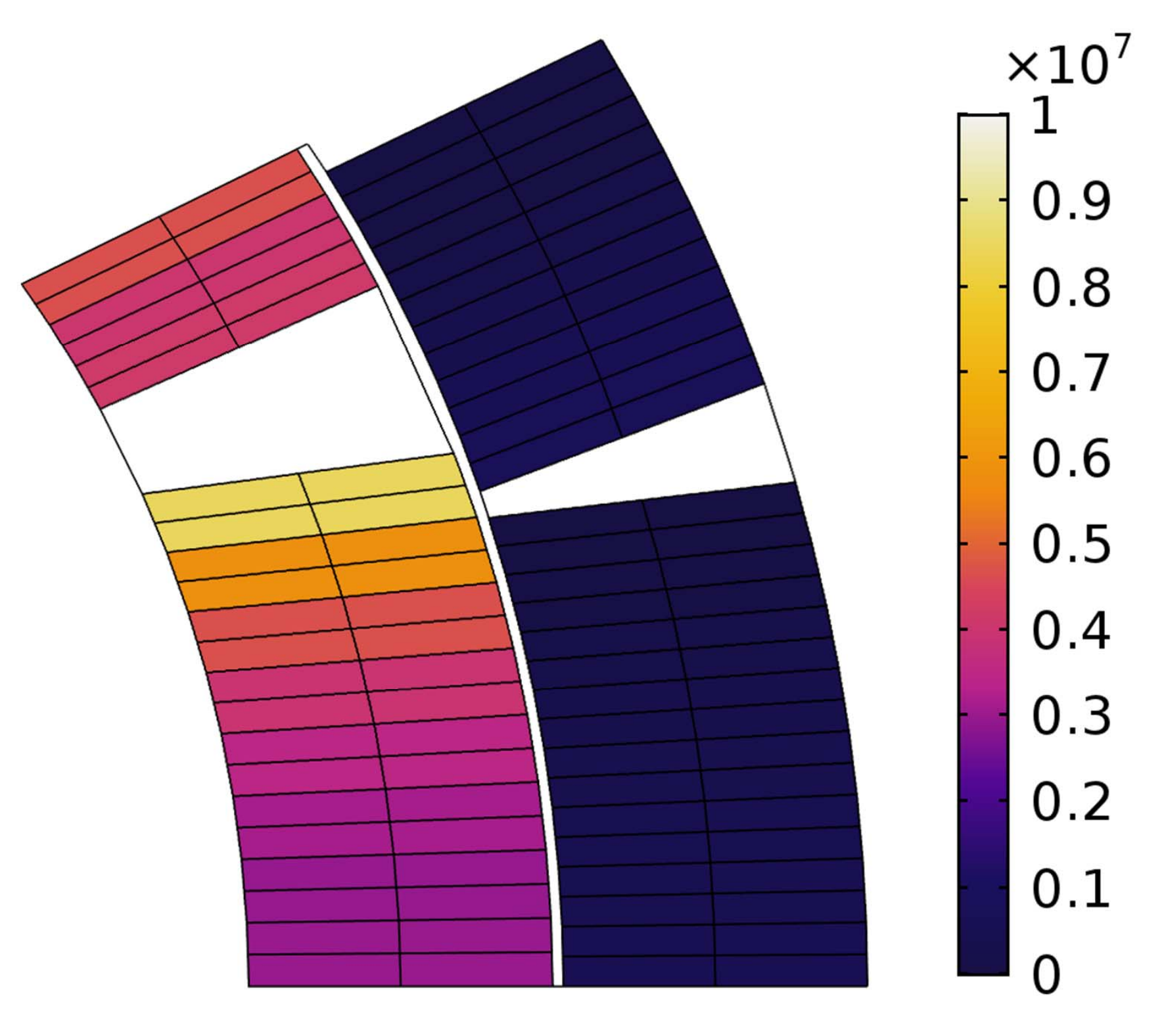}
   		\caption{}
   		\label{fig:No2}
\end{subfigure}
\begin{subfigure}[b]{0.23\textwidth}
   		\centering
   		\includegraphics[width=4.4cm]{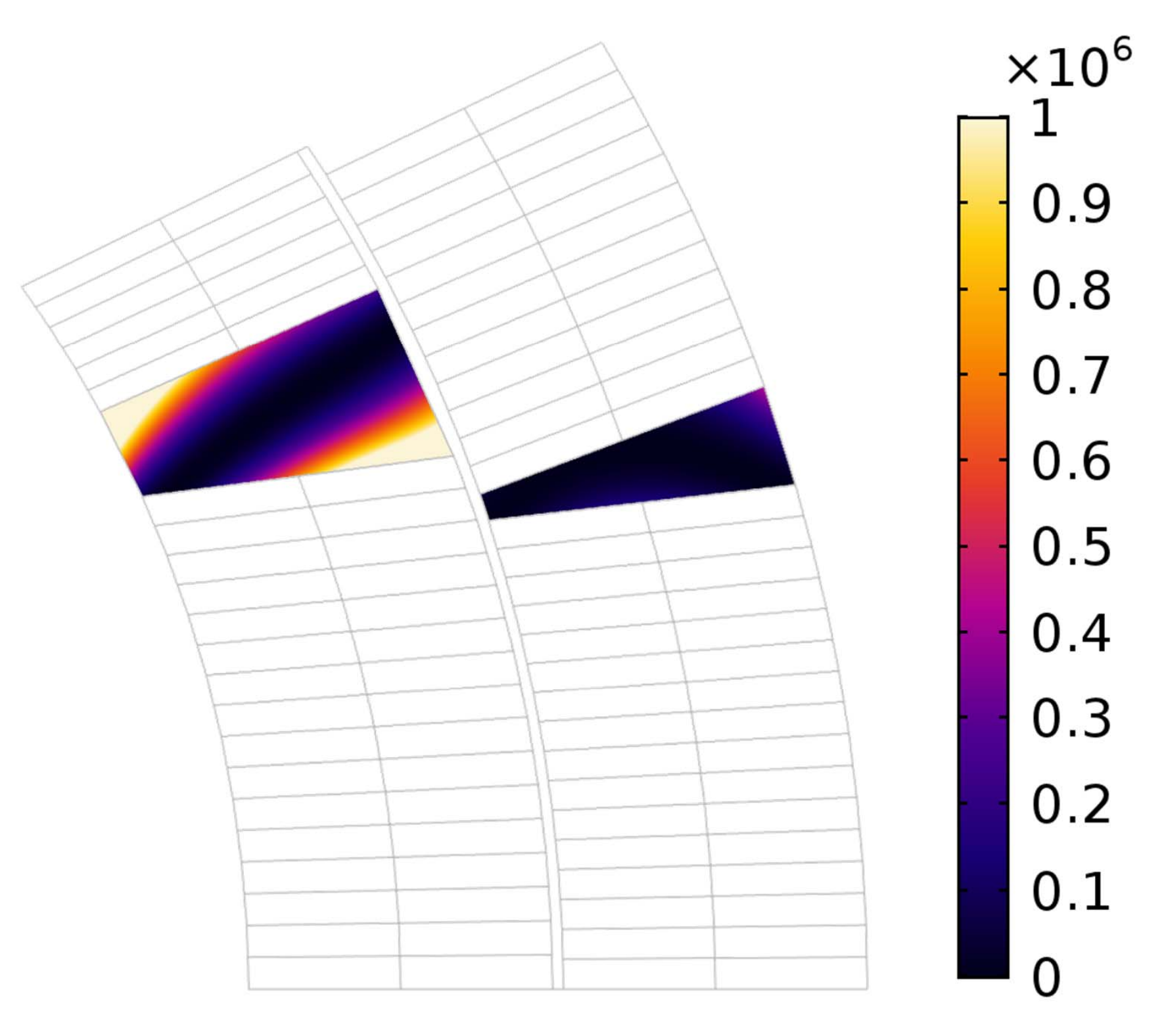}
   		\caption{}
   		\label{fig:No3}
\end{subfigure}
   \begin{subfigure}[b]{0.23\textwidth}
   		\centering
   		\includegraphics[width=4.4cm]{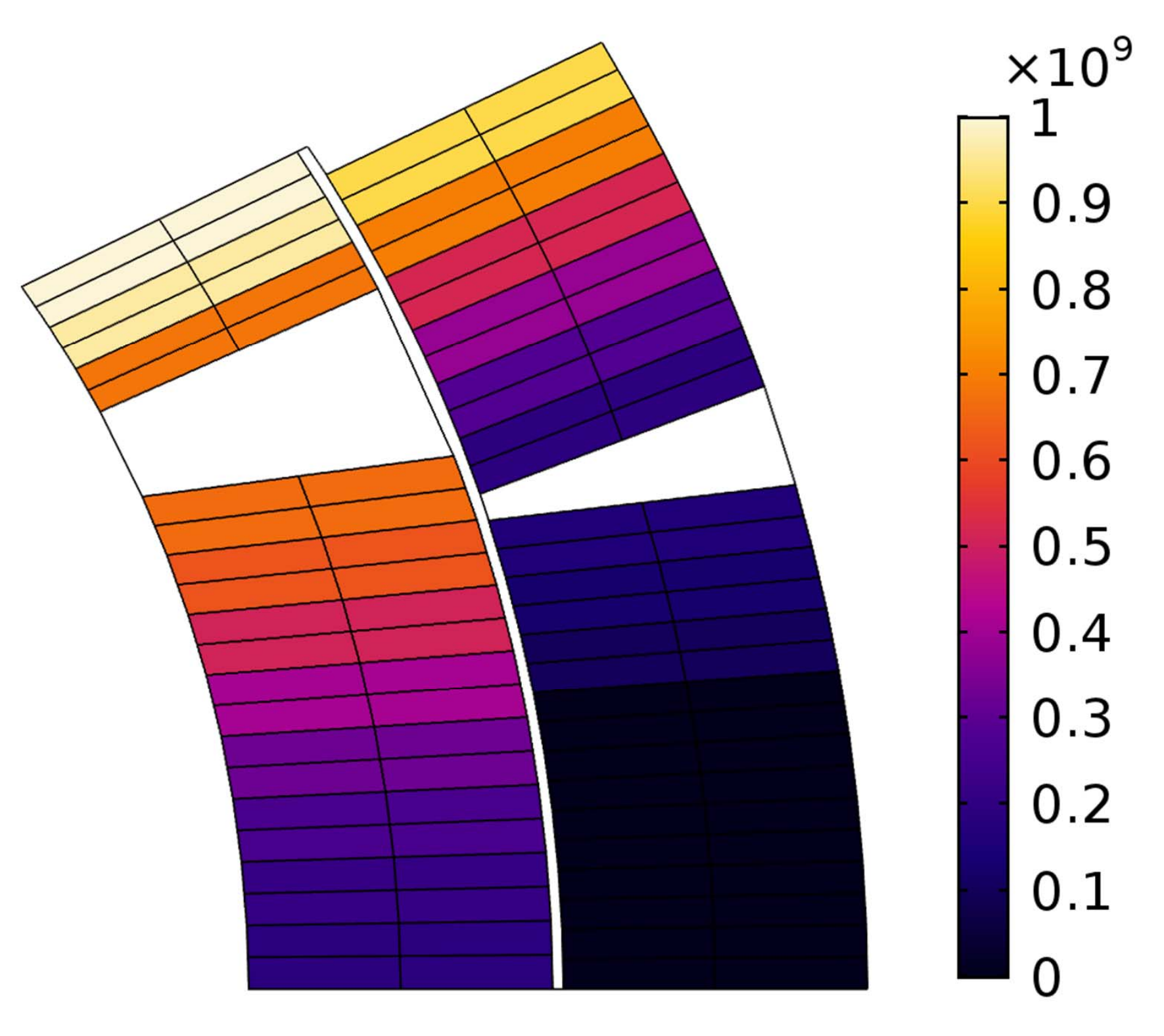}
   		\caption{}
   		\label{fig:No4} 
\end{subfigure}
\begin{subfigure}[b]{0.23\textwidth}
   		\centering
   		\includegraphics[width=4.4cm]{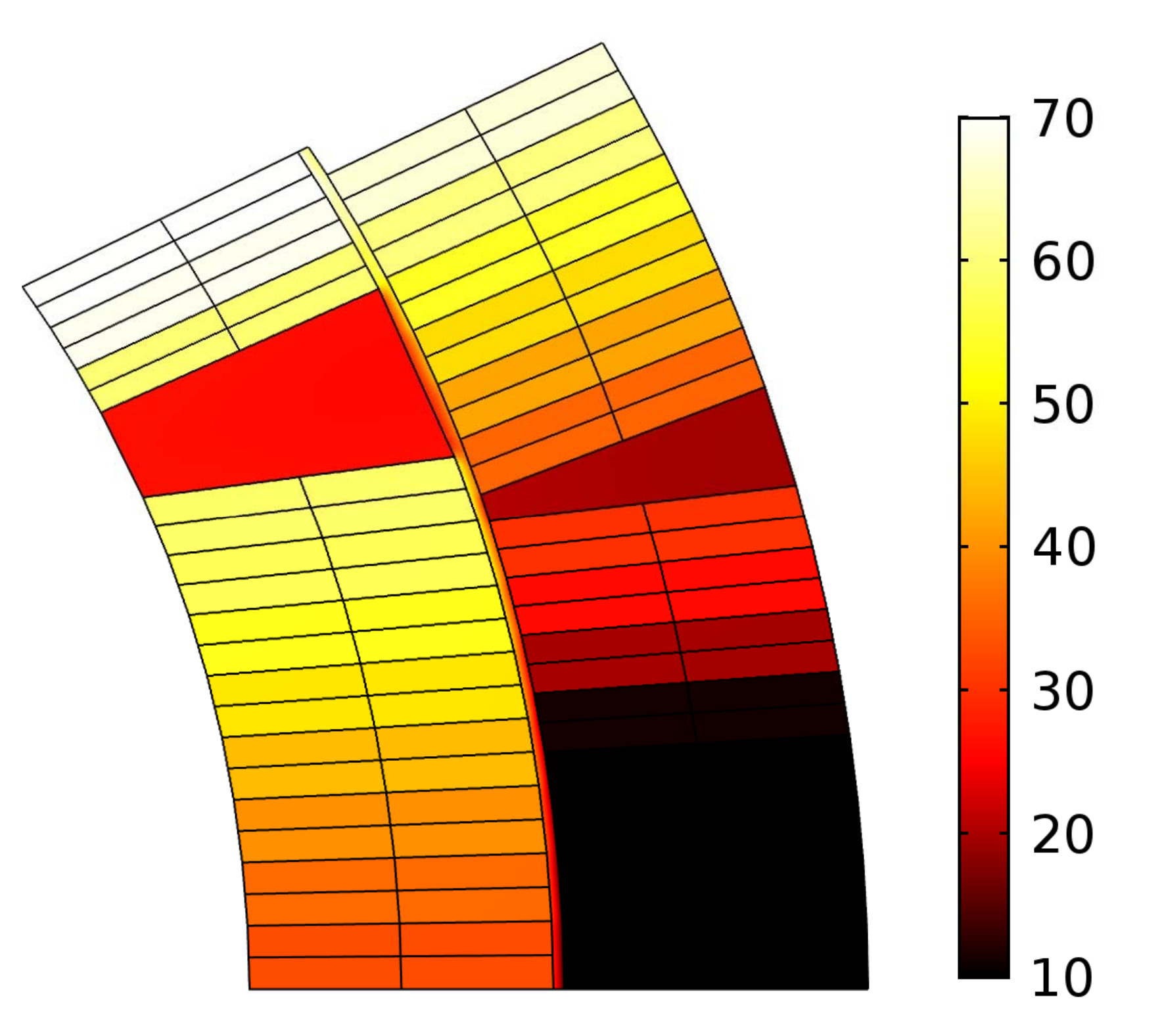}
   		\caption{}
   		\label{fig:No5}
\end{subfigure}
\begin{subfigure}[b]{0.23\textwidth}
   		\centering
   		\includegraphics[width=4.4cm]{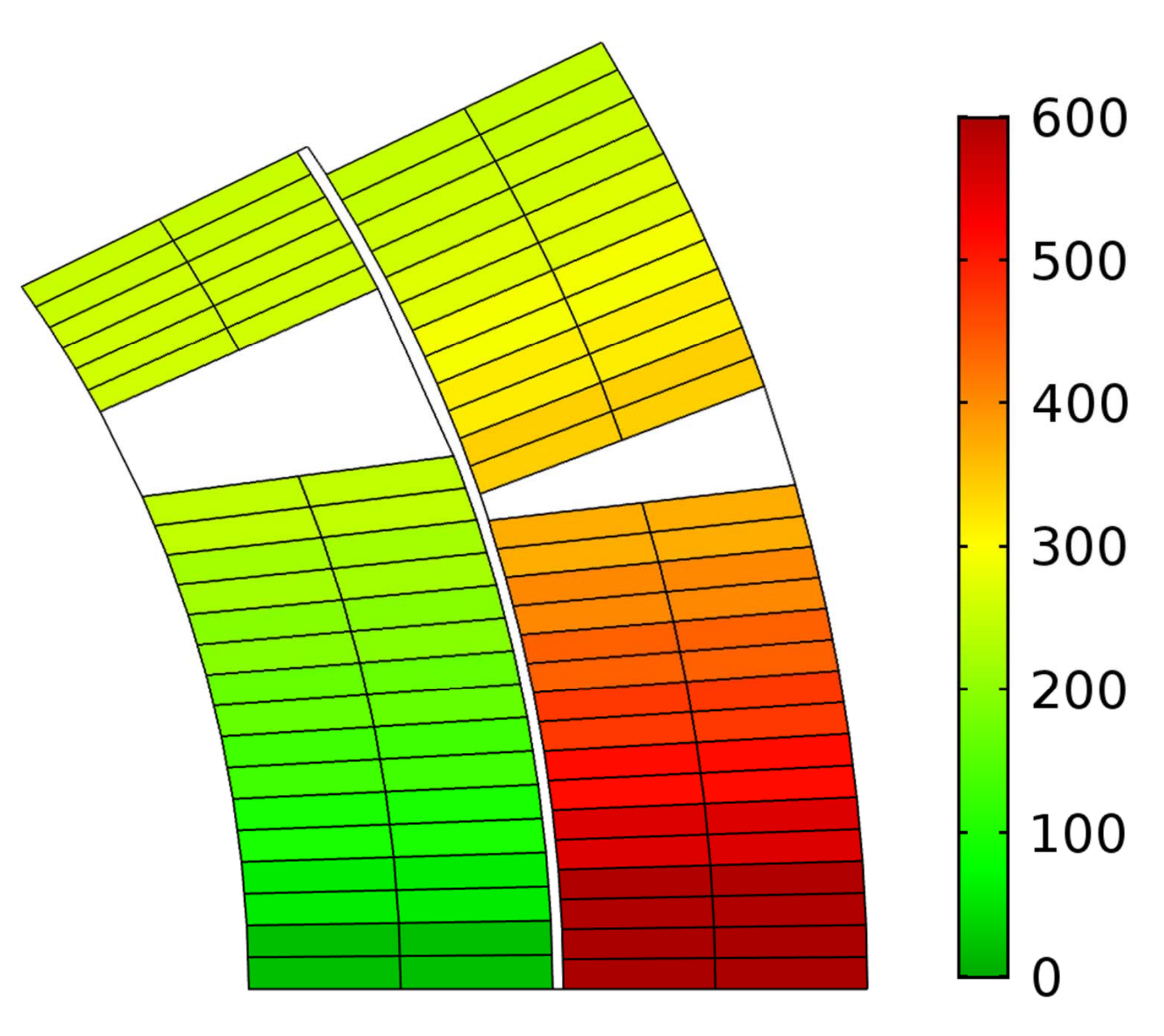}
   		\caption{}
   		\label{fig:No6}
\end{subfigure}
\caption{Co-simulation results, at $50~\mathrm{ms}$. (a) Inter-filament coupling losses, in $\mathrm{W/m^3}$. (b) Inter-strand coupling losses, in $\mathrm{W/m^3}$. (c) Eddy-current losses, in $\mathrm{W/m^3}$. (d) Ohmic losses, in $\mathrm{W/m^3}$. (e) Temperature map, in $\mathrm{K}$. (f) Voltage-to-ground map, in $\mathrm{V}$. }
\label{Meq_IF_IS_CC}
\end{figure}
The proposed formulation is applied to simulate the Hi-Lumi MQXF quadrupole magnet~\cite{ferracin2016development} in a circuit, as a first application to magnets based on $\mathrm{Nb_{3}Sn}$ superconducting technology. 
The two-port component interface (Sec.~\ref{Interface}), combined with the equivalent lumped-parameter representation proposed in~\cite{garcia2017optimized}, is used to interface the FEM model with a PSPICE circuit. The chosen value for the circuital inductance of the magnet is $60~\text{mH}$~\cite{garcia2017optimized}.
The circuit simulates a test bench which is arranged as in~\cite{bortot2016consistent}, where the formulation has been cross-checked for Nb-Ti magnets. The resistor in parallel with the magnet is equal to  ${175~\text{m}\Omega}$. 

The FEM and the circuit models are co-simulated applying the waveform relaxation technique~\cite{schops2010cosimulation}.  
The results refer to a simulation scenario called quench-back. The magnet's nominal operating condition at 1.9~K, 17.8~kA, determines the initial condition of the circuit. Then, the current dynamics and the subsequent field variation induce losses in the coil assembly due to the IFCCs (Fig.~\ref{fig:No1}) and ISCCs (Fig.~\ref{fig:No2}) in the cable, and the eddy currents in the wedges (Fig.~\ref{fig:No3}). Dynamic losses  influence the current decay, contributing to dissipate the stored energy, and heat up the coil until superconductivity is lost. While the quench increases the magnet's temperature, the coil resistance contributes to an even faster discharge of the current, limiting the deposition of the Ohmic losses (Fig.~\ref{fig:No4}) which determine the temperature distribution (Fig.~\ref{fig:No5}).
The combination of the resistive and inductive voltage leads to the voltage-to-ground profile presented in Fig.~\ref{fig:No6}. The model is meshed with $13\cdot10^3$, $2^{\mathrm{nd}}$-order elements, over which the the monolithically-coupled field equations are discretized and processed by the direct solver PARDISO. The solution required 190 time steps, using a $2^{\mathrm{nd}}$-order backward differentiation formula. The computational time for a magnet's quadrant discharge (500 $\text{ms}$) is about $2\mathrm{h}$, on a standard workstation.


\section{Conclusions} \label{Conclusion}
We successfully developed a 2-D finite-element electro-thermal model for accelerator magnets. The model accounts for the dynamic effects occurring both in the superconducting cable through equivalent magnetization formulations, and in the copper wedges. The thermal formulation includes the coil's structural elements and accounts for the layer-to-layer propagation. A two-port component interface allows the model to be co-simulated with an external circuit. Assuming the need of a fast extraction of the stored energy, the quench-back scenario is analyzed, and the magnet's integrity is assessed.
  

\section{Acknowledgement} \label{Acknowledgement} 
The authors would like to thank K. Kr{\'o}l, and J.C. Garnier from CERN for the help provided in the coding of the FEM workflow in Java, and Dr. Sven Friedel and COMSOL Switzerland for continued support and valuable suggestions during the development of the project.

  

\end{document}